\documentclass{Pos}
\usepackage{epsf}
\usepackage{cite}
\usepackage{amsmath}
\usepackage{wrapfig}

\def\beq{\begin{equation}}
\def\eeq{\end{equation}}
\def\bea{\begin{eqnarray}}
\def\eea{\end{eqnarray}}
\def\beqa{\begin{equation}\begin{array}{l}}
\def\eeqa{\end{array}\end{equation}}
\def\eqlab#1{\label{eq:#1}}
\def\figlab#1{\label{fig:#1}}

\def\Figref#1{Fig.~\ref{fig:#1}}

\def\la{\lambda} \def\La{{\Lambda}}

\def\dd{{\rm d}}

\def\nn{\nonumber}

\def\im{\mbox{Im}}

\newcommand{\nc}{\newcommand}
\nc{\lb}{\langle}
\nc{\rk}{\rangle}
\nc{\Blb}{\Big\langle}
\nc{\Brk}{\Big\rangle}

\nc{\mi}{\!\!\mid\!\!}
\nc{\Ra}{\Rightarrow}
\nc {\cd}{\partial}
\nc {\sla}{\slashed}
\nc{\ro}{\mathrm}
\nc{\ca}{\mathcal}
\nc{\sr}{\mathscr}
\nc{\Tr}{\ro{Tr}}
\nc{\Str}{\ro{Str}}
\nc{\realtrace}{\ro{Re\; Tr}}
\nc{\maxrealtrace}{\ro{max\, Re\; Tr}}
\nc{\ud}{\ro{d}}

\nc{\cLe}{\ca{L}_{\ro{eff}}}
\nc {\ti}{\tilde}
\nc{\f}{\frac}
\nc{\da}{\dagger}
\nc{\SU}{\ro{SU}}

\nc{\darrow}{\stackrel{\leftrightarrow}{\cd}}
\nc{\darrows}{\stackrel{\leftrightarrow}{\sla{\cd}}}
\nc{\Darrows}{\stackrel{\leftrightarrow}{\sla{D}}}
\nc {\mpisq}{m_{\pi}^2}
\nc{\Mc}{\stackrel{\circ}{M}}
\nc{\cc}{\stackrel{\circ}{c}}
\nc{\kc}{\stackrel{\circ}{\kappa}}
\nc{\McHB}{\stackrel{\circ}{M}_N^{\ro{HB}}}
\nc{\McB}{\stackrel{\circ}{M}_N^{\ro{B}}}
\nc{\ccHB}{\stackrel{\circ}{c}_1^{\ro{HB}}}
\nc{\ccB}{\stackrel{\circ}{c}_1^{\ro{B}}}
\nc{\kcHB}{\stackrel{\circ}{\kappa}_{\ro{isov}}^{\ro{HB}}}
\nc{\kcB}{\stackrel{\circ}{\kappa}_{\ro{isov}}^{\ro{B}}}
\nc{\gc}{\stackrel{\circ}{g}}
\nc{\fc}{\stackrel{\circ}{f}}
\nc{\hM}{\hat{M}_N}
\nc {\eqb}{\begin{equation}}
\nc {\eqe}{\end{equation}}
\nc {\eqab}{\begin{eqnarray}}
\nc {\eqae}{\end{eqnarray}}

\title{Pion-mass dispersion relation in the baryon sector }

\ShortTitle{Pion-mass dispersion relation }

\author{\speaker{Vladimir Pascalutsa}\thanks{Supported by the Deutsche Forschungsgemeinschaft (DFG) through Collaborative 
Research Center SFB 1044.}, \ Marc Vanderhaeghen 
       \\
       Institut f\"ur Kernphysik, Johannes Gutenberg Universit\"at, Mainz D-55099, Germany
       }

\author{Jonathan M. M. Hall\\
CSSM, School of Chemistry and Physics, University of Adelaide 5005,
  Australia}
  
 \author{Tim Ledwig\\
Departamento de F\'\i sica Te\'orica and IFIC, Centro Mixto
Universidad de Valencia-CSIC, E-46071 Valencia, Spain}

\abstract{By looking at the complex plane of the pion-mass squared
we establish a dispersion relation which the static quantities, such 
as baryon masses, magnetic moments, polarizabilities, should obey.
This dispersion relation yields insight into the differences 
between the heavy-baryon and relativistic calculations
in the baryon sector of chiral perturbation theory.}

\FullConference{7th International Workshop on Chiral Dynamics\\
		August 6 -10, 2012\\
		Jefferson Lab, Newport News, Virginia, USA }

\begin{document}

\section{Introduction}

It is well-known that the chiral perturbation theory ($\chi$PT) 
is able to predict 
some `non-analytic' dependencies of static quantities (masses, magnetic moments, etc.)
on pion-mass squared, or the quark mass ($m_\pi^2 \sim m_q$).
It is therefore interesting to examine the origin of these dependencies arising
by considering the  analytic properties of chiral expansion in the entire 
complex $m_\pi^2$ plane. 
\begin{figure}[h]
\centerline{\epsfclipon   \epsfxsize=11.5cm%
  \epsffile{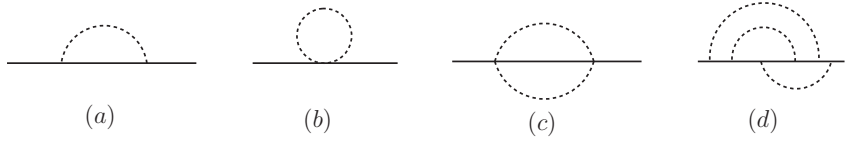} 
}
\caption{Examples of chiral-loop corrections to the nucleon mass. Nucleon (pion) propagators are denoted by solid (dashed) lines.}
\figlab{nuclse}
\end{figure}

\begin{wrapfigure}{r}{2.2in}
\vspace{-4mm}
\centerline{\epsfclipon   \epsfxsize=4.cm%
  \epsffile{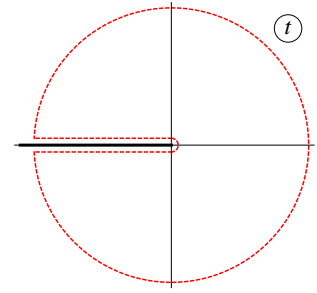} 
}
\caption{The branch cut and the contour defining the analyticity domain
in the complex plane of 
$t=m_\pi^2$.}
\figlab{cont1}
\end{wrapfigure}
Considering the chiral loops 
with external nucleons on shell, as in the graphs of 
\Figref{nuclse}   arising in the original 
(manifestly Lorentz-invariant) formulation of baryon $\chi PT$ (B$\chi PT$)
\cite{GSS89},
we observe that they are analytic functions of $m_\pi^2$ everywhere
except for the branch cut along the negative axis, see \Figref{cont1}.
In this case it is possible to write down a simple dispersion relation
in the pion-mass squared~\cite{Ledwig:2010nm}:
\beq
\eqlab{disprel}
 f(m_\pi^2) = -\frac{1}{\pi}\int\limits_{-\infty}^0 \dd t \, \frac{\im\, f(t)}{t-m_\pi^2+i0^{+}}\,,
\eeq
where $f$ is a chiral-loop correction to a static quantity, or the static quantity
itself; $0^+$ is an infinitesimal positive number.

Several applications of this dispersion relation have been discussed
in \cite{Ledwig:2010nm}. Here we shall focus on
 a study of large discrepancies between the leading-order 
 heavy-baryon (HB) \cite{JeM91a,Bernard:1995dp}
and B$\chi$PT~\cite{GSS89,Fuchs:2003qc}  calculations
encountered in e.g.\ Refs.~\cite{Pascalutsa:2004ga,Geng:2008mf,Lensky:2009uv,Strikman:2010pu}.

\section{B$\chi$PT vs.\ HB$\chi$PT at $\ca{O}(p^3)$}

The chiral expansion of a static quantity $f$ 
is an expansion in the quark mass 
$m_q$ around the chiral limit ($m_q~\to~0$), which in $\chi$PT becomes
an expansion in $p=m_\pi/\La_\chi$, the mass of the 
pseudo-Goldstone boson of spontaneous chiral 
symmetry breaking over the scale of chiral symmetry breaking 
$\La_\chi \simeq 4\pi f_\pi \approx 1$ GeV.
Because of the branch cut in the complex-$m_\pi^2$ plane 
along the negative real-axis, the chiral expansion is not a series
expansion (
otherwise, it would have a zero radius of convergence), but rather 
an expansion in non-integer powers of $m_\pi^2 \propto m_q$.

By writing the dispersion integral as:
\eqb
\label{eqn:disprel5}
f(m_\pi^2) = -\f{1}{\pi}\left( \, \int\limits_{-\La_\chi^2}^0 
+ \int\limits_{-\infty}^{-\La_\chi^2}\, \right) \ro{d}t \,
\f{\ro{Im}\, f(t)}{t-m_\pi^2} ,
\eqe
it is evident that the second integral can  be expanded in
integer powers of $
m_\pi^2/\La_\chi^2
$. Hence this term is of analytic form and can only affect the values of the
LECs. Indeed, the physics above the scale $\La_\chi$ is not
described by $\chi$PT and therefore its effect should be absorbable 
in the LECs.

\begin{figure}[b]
\begin{minipage}[c]{.52\linewidth}
\begin{center}
\includegraphics[height=1.65\hsize]{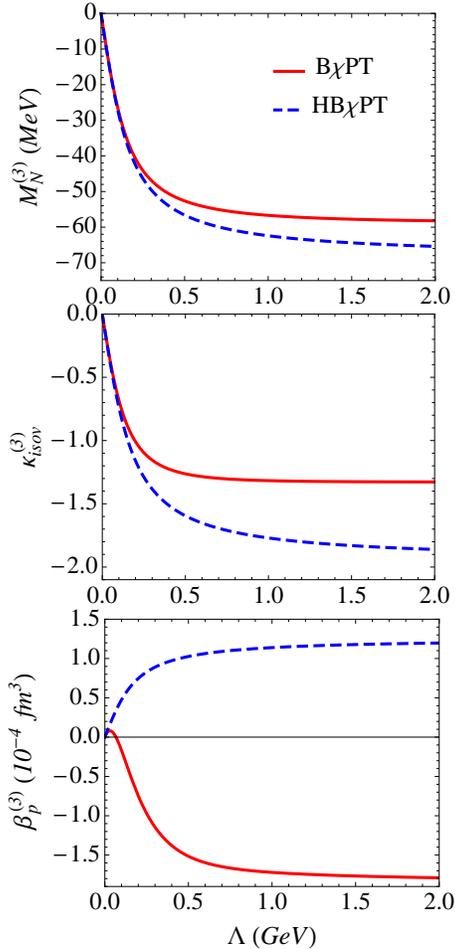}
\end{center}
\end{minipage}
\hspace{.06\linewidth}%
\begin{minipage}[c]{.35\linewidth}
\caption{
 The cutoff-dependence of leading-order loop 
contributions to various nucleon quantities (mass, isovector AMM, and
proton's magnetic polarizability) calculated in HB$\chi$PT 
(blue dashed curves) and B$\chi$PT (red solid curves). 
}
\label{fig:LaDep}
\end{minipage}
\end{figure}

The second integral generates an infinite number
of analytic terms, while the number of LECs to a given order of the calculation 
is finite. The higher-order analytic terms are present and not compensated
by the LECs at this order, but their effect should not exceed the uncertainty 
in the calculation due to the neglect of all the other higher-order terms.
That is, the second integral can be dropped, while the resulting
cutoff-dependence represents the uncertainty due to higher-order effects.
We are thus led to examine the cutoff dependence of the pion-mass
dispersion relation \cite{Hall:2012iw}:
\eqb
\label{eqn:disprel3}
 f(m_\pi^2; \La^2) = -\f{1}{\pi}\!\int\limits_{-\La^2}^0\!\!\ro{d}t \,
\f{\ro{Im} \, f(t)}{t-m_\pi^2} \left(\f{m_\pi^2}{t}\right)^n,
\eqe
 where $n$ indicates the number of subtractions around the chiral limit.
Our main aim is to see at which values of 
the cutoff any deviation occurs between the 
HB- and B$\chi$PT results.  
If the deviation begins at $\La \ll 1$ GeV, then the differences between 
the two expansions cannot be reconciled in a natural way. 
In the next section, this situation is examined using 
several specific examples, and for each of them a different picture is 
obtained (cf.~Fig.~\ref{fig:LaDep}).

 \label{sect:calcs}
 
At chiral order $p^3$, the imaginary parts of the nucleon mass,
the proton
and neutron AMMs, and the magnetic polarizability of the 
proton are given by:
\begin{subequations}
\begin{align}
\label{ImMN}
\mathrm{Im}\, M_N^{(3)} (t) & =  
 \frac{3 g_A^2 \hat M_N^3}{(4\pi f_\pi)^2} \frac{\pi \tau}{2} 
\Big( \f{1}{2} \tau +\la \Big)\, \theta(-t)\,,\\
\label{eqn:Imkp}
\ro{Im}\, \kappa_p^{(3)} (t)&=
\frac{g_A^2 \hat M_N^2 }{(4\pi f_\pi)^2}  \frac{2 \pi  }{\la}  \Big(\f{1}{2}\tau
 + \la \Big)^2
\Big[ 1-\f{3}{2} \Big( \f{1}{2} \tau +\la \Big) \Big]\, \times \theta(-t)\, 
, \\
\label{eqn:Imkn}
\ro{Im}\, \kappa_n^{(3)} (t)&=
-\frac{g_A^2 \hat M_N^2 }{(4 \pi f_\pi)^2}  \frac{2 \pi  }{ \la}  \Big(\f{1}{2}
 \tau + \la \Big)^2\, \theta(-t)\,, \\
\mathrm{Im}\, \beta^{(3)}_p (t) & = - 
\frac{(e^2/4\pi) \, g_A^2}{(4 \pi f_\pi)^2 \hat M_N }
\frac{ \pi \tau }{24\la^3} \Big[ 2-72 \la + (418\la-246) \,\tau \nn\\
& - (316\la-471) \,\tau^2 
 +(54\la-212)\,\tau^3 +27\tau^4\Big] \,\theta(-t),
 \end{align}
  \end{subequations}
 where $\hat M_N\simeq 939$ 
 MeV is the physical nucleon mass,  $e^2/4\pi \simeq 1/137$ is 
 the fine-structure constant, and
 the following dimensionless variables are introduced:
 \eqb
 \tau = \f{t}{\hat M_N^2}, \quad \la = \sqrt{\f{1}{4} \tau^2 -\tau}\,.
 \eqe
  
  The corresponding HB expressions at order $p^3$ can be
   obtained by keeping only the leading in $1/\hat M_N$ term 
(i.e, $\la \approx \sqrt{-\tau}$, etc.):
 \begin{subequations}
\begin{align}
\mathrm{Im}\, M_N^{(3)} (t) & \stackrel{\ro{HB}}{=}  
 \frac{3 g_A^2 \hat M_N^3}{(4\pi f_\pi)^2} \frac{\pi \tau}{2} 
\sqrt{-\tau} \, \theta(-t)\,,\\
\label{eqn:HBImkp}
\ro{Im}\, \kappa_p^{(3)} (t)& \stackrel{\ro{HB}}{=}
\frac{ g_A^2 \hat M_N^2 }{(4\pi f_\pi)^2}\, 2 \pi \sqrt{-\tau}
\, \theta(-t) \stackrel{\ro{HB}}{=} - \, \ro{Im}\, \kappa_n^{(3)} (t), \\
\mathrm{Im}\, \beta^{(3)}_p (t) & \stackrel{\ro{HB}}{=} 
\frac{(e^2/4\pi) \, g_A^2}{(4 \pi f_\pi)^2 \hat M_N }
\frac{ \pi }{12\sqrt{-\tau}}  \,\theta(-t).
 \end{align}
 \label{eqn:HBImMN}
\end{subequations}

The full, renormalized result for a given quantity is obtained by substituting
these imaginary parts into the dispersion relation of Eq.~(\ref{eqn:disprel3}). 
The number of subtractions required in each case differ: $n=2$ for $M_N$,
$n=1$ for AMMs, and no subtractions for polarizability.  
The resulting expressions can be found in \cite{Hall:2012iw}.

The heavy-baryon expressions can be
obtained by picking out the leading in $1/\hat M_N$ term, or equivalently, 
by substituting the corresponding imaginary parts from Eq.~(\ref{eqn:HBImMN}),
into the dispersion relation. In the latter case, 
the same integral is encountered in all of the examples:
\begin{align}
J(m_\pi; \La)\equiv \int\limits_{-\La^2}^0\!\ro{d}t \, \f{1}{(t-m_\pi^2)\sqrt{-t}}
 =  -\f{2}{m_\pi}\arctan\f{\La}{m_\pi}.
\end{align}
All of the above quantities to $\ca{O}(p^3)$ in HB$\chi$PT are given by this 
integral, up to an overall constant, and a factor of $m_\pi^{2n}$. 
$n$ is the number of subtractions (or pertinent LECs) at this order.  
%
In Fig.~\ref{fig:LaDep}, the resulting cutoff-dependence of 
the above loop contributions is shown at the physical value of the pion mass: 
$m_\pi \simeq 139$ MeV. Each quantity (mass, isovector AMM, and 
polarizability) is presented in a separate
panel, where the results 
within the relativistic ChPT and using the HB expansion are displayed.

The figure illustrates 
the following two features: 
\begin{enumerate}
\item  The HB$\chi$PT results have a stronger cutoff-dependence than 
the B$\chi$PT results, indicating a larger impact of the unknown 
high-energy physics to be renormalized by higher-order LECs.
Quantitatively, the residual cutoff-dependence in HB$\chi$PT 
falls as $1/\La$ in all of the considered examples, while 
the dependence in the case of B$\chi$PT behaves as 
$ 1/\La^2$ for $M_N$, and as  $ 1/\La^4$ for both AMMs and $\beta_p$.
\item The  HB- and B$\chi$PT results are guaranteed to be the same at 
small values of $\La$, as can be seen by taking derivatives of 
Eq.~(\ref{eqn:disprel3}) 
with respect to $\La^2$, at $\La=0$. However, at finite values of 
$\La$ the differences
are appreciable. Observing significant differences for $\La$ of order $m_\pi$, 
as in the case of $\beta_p$, 
indicates that the size of the $1/\hM$ terms 
is largely underestimated in HB$\chi$PT.
\end{enumerate}

\section{Conclusion and outlook}
\label{sect:conc}
The HB$\chi$PT  and B$\chi$PT can be viewed as two different ways 
of organizing the chiral EFT expansion in the baryon sector. While 
the heavy-baryon expansion
is often considered to be more consistent from the power-counting point of 
view, it appears to be less natural. 
Certain terms that are nominally 
suppressed by powers of $m_\pi/M_N$, and hence dropped in HB$\chi$PT as being `higher order', appear to be significant in explicit calculations.   

The problem is more pronounced in some quantities and less in others.
To quantify this, one needs to note the power
of the expansion parameter at which the chiral loops begin to contribute
to the quantity in question.
For the considered examples of the nucleon mass, AMMs, and polarizability, 
this power index is $3$, $1$, and $-1$, respectively. 
The smaller the index, the
greater is the 
 difficulty for HB$\chi$PT to describe this quantity in a natural way. 

The negative index simply means that the chiral expansion of that quantity 
begins with negative powers of $m_\pi$. Apart from polarizabilities, 
the most notable quantities of this kind are the coefficients of the 
effective-range expansion of the nuclear force. The non-relativistic
$\chi$PT in the two-nucleon sector \cite{Kaplan:1998we} 
failed to describe these 
quantities \cite{Cohen:1998jr}, 
thus precluding the idea of `perturbative pions' in this sector. 
The present work encourages us to think that B$\chi$PT can 
solve this problem in a way similar to the case of 
nucleon polarizabilities \cite{Lensky:2009uv}.

\end{document}